\documentclass[reprint,notitlepage,showpacs,aps,prd,nofootinbib,superscriptaddress,longbibliography]{revtex4-2}

\usepackage{graphicx}
\usepackage[utf8]{inputenc} 
\usepackage{placeins}
\usepackage{amsmath}
\usepackage{amssymb}
\usepackage{float}
\usepackage{xcolor,color,colortbl}
\usepackage{comment}
\usepackage{multirow}
\usepackage{color}
\usepackage{soul}
\usepackage{ulem}

\def\beq{\begin{equation}}
\def\eeq{\end{equation}}
\def\bea{\begin{eqnarray}}
\def\eea{\end{eqnarray}}

\usepackage{amsbsy}
\usepackage{bm}
\usepackage{color}
\usepackage{fancyhdr}
\usepackage[pdftex]{epsfig}
\usepackage[colorlinks=true,linkcolor=blue,citecolor=purple,urlcolor=magenta,filecolor=blue]{hyperref}
\usepackage{slashed}

\begin{document}

\bigskip
\vskip 6ex

\author{J. M. Cabarcas}
\email{josecabarcas@usta.edu.co}
\affiliation{Universidad Santo Tomás, Colombia}

\author{Alexander Parada}
\email{alexander.parada@esap.edu.co}
\affiliation{Escuela Superior de Administración Pública, Territorial Meta, Carrera 31A No.34A-23, C\'{o}digo Postal 500001, Villavicencio, Colombia}

\author{N\'{e}stor Quintero-Poveda}
\email{nquintero@ut.edu.co}
\affiliation{Departamento de F\'{i}sica, Universidad del Tolima, C\'{o}digo Postal 730006299, Ibagu\'{e}, Colombia}
\affiliation{Facultad de Ciencias B\'{a}sicas, Universidad Santiago de Cali, Campus Pampalinda, Calle 5 No.62-00, C\'{o}digo Postal 76001, Santiago de Cali, Colombia}

\vspace{2cm}

\title{Implications of NSI constraints from ANTARES and IceCube on a simplified $Z^\prime$ model}

\bigskip

\begin{abstract}
Recently the neutrino experiments ANTARES and IceCube have released new constraints to the non-standard neutrino interaction (NSI) parameter $\epsilon^d_{\mu\tau}$ (flavor off-diagonal). These new constraints are stronger than those obtained from a combination of COHERENT and neutrino oscillation data. In the light of the recent constraints from ANTARES and IceCube data on the NSI parameter $\epsilon^d_{\mu\tau}$, in this work, we study the new physics implications on the parameter space of a simplified $Z^\prime$ model with lepton flavor violating ($\mu\tau$) couplings. For a $Z^\prime$ boson with a mass heavier than the $\tau$ lepton, our results show that ANTARES and IceCube can provide additional constraints to such a new physics scenario with $\mu\tau$ couplings, when compared to bounds from low-energy flavor physics. Moreover, these neutrino experiments can exclude a similar region than ATLAS experiment, showing the potential to provide complementary information to the one obtained from direct searches at the Large Hadron Collider. The impact of the expected sensitivity at DUNE and Belle II experiments is also studied.
\end{abstract}

\maketitle

\section{Introduction}

We have known for many years that neutrinos are massive particles, thanks to neutrino oscillation mechanism, successfully demonstrated by a large number of precision neutrino physics experiments. However, the Standard Model (SM) does not provide an explanation for the existence of the neutrino mass. Efforts to describe the origin of this neutrino property have showed the way towards proposing new couplings (between neutrinos and unidentified particles) known as non-standard interactions (NSI) \cite{Esteban:2018ppq,Escrihuela:2011cf,Davidson:2003ha,Coloma:2019mbs,Ohlsson:2012kf,Miranda:2015dra}.
These new interactions including neutral currents (NC) can be described through the general Lagrangian
\begin{equation}
\mathcal{L}_{\rm NSI} = - 2\sqrt{2} G_F \ \epsilon^{f X}_{\alpha\beta} (\bar{\nu}_\alpha \gamma^\rho P_L \nu_\beta)(\bar{f} \gamma_\rho P_X f) ,
\end{equation}

\noindent where $G_F$ is the Fermi constant, $f=e,u,d$ are the first-family fermions, $X = L, R$ represents the chirality projection $P_{L,R} = (1 \mp \gamma_5)/2$, and $\alpha,\beta = e, \mu,\tau$ is the neutrino flavor index. All the new physics (NP) effects are codified through the dimensionless parameters $\epsilon^{f X}_{\alpha\beta}$. 
The NSI parameters can be flavor-preserving $\alpha = \beta$ (diagonal) and/or flavor-violating $\alpha \neq \beta$ (off-diagonal).
A considerable amount of constraints on these parameters have been obtained from different sources, mostly including data of neutrino oscillations experiments. Particularly, in the off-diagonal NSI $\mu-\tau$ sector with down-quarks ($\epsilon^{d}_{\mu\tau}$), we can find in the literature limits on this NSI parameter of the order $-0.012<\epsilon^{d}_{\mu\tau}<0.009$~\cite{Esteban:2018ppq}  and $-0.007<\epsilon^{d}_{\mu\tau}<0.007$~\cite{Escrihuela:2011cf}. Moreover, analysis from COHERENT experiment data led to the constraint $\epsilon^{d}_{\mu\tau}<0.20$ \cite{Giunti:2019xpr} at 90\% of confidence level, and a combination between neutrino oscillation and COHERENT data has allowed a better result on this parameter, $-0.008<\epsilon^{d}_{\mu\tau}<0.008$ \cite{Coloma:2019mbs}. Neutrinos telescopes such as ANTARES~\cite{ANTARES:2021crm} and IceCube~\cite{IceCube:2022ubv} are also committed to a exhaustive searching for NSI. This kind of experiments  registered the observation of atmospheric muon neutrinos, which can oscillate mainly to tau neutrinos, therefore there is a high sensitivity to the $\epsilon^d_{\mu\tau}$ parameter, allowing them to report the most stringent experimental constraints on NSI parameter $\epsilon^d_{\mu\tau}$ to date. In the case of ANTARES~\cite{ANTARES:2021crm}, the analysis correspond to ten years of data collected from 2007 to 2016, in the range of energies from 16 GeV to 100 GeV to give rise to the constraints
\begin{equation}
    \text{ANTARES:} 
\begin{cases} 
-4.7\times 10^{-3}<\epsilon^{d}_{\mu\tau}<2.9\times 10^{-3}\quad\text{(NO)}, \\
-2.9\times 10^{-3}<\epsilon^{d}_{\mu\tau}<4.7\times 10^{-3}\quad\text{(IO)},
\end{cases}  
\end{equation}

\noindent at 90$\%$ of confidence level and considering normal (NO) and inverted (IO) neutrino mass ordering, respectively~\cite{ANTARES:2021crm}. As concerns IceCube Collaboration~\cite{IceCube:2022ubv}, this experiment performed an analysis using a sample of 305,735 muon tracks from neutrino charged current interactions observed from May of 2011 to May of 2019, reporting the following limits,
\begin{equation}
    \text{IceCube:} 
\begin{cases} 
-4.1\times 10^{-3}<\epsilon^{d}_{\mu\tau}<3.1\times 10^{-3}\quad\text{(NO)}, \\
-3.0\times 10^{-3}<\epsilon^{d}_{\mu\tau}<4.0\times 10^{-3}\quad\text{(IO)},
\end{cases}  
\end{equation}
at 90$\%$ of confidence level~\cite{IceCube:2022ubv}.  \medskip

Furthermore, there are high expectations about the sensitivity of future long-baseline neutrino oscillation experiments to the NSI, namely, DUNE, NO$\nu$A, T2K, T2HK, among others \cite{Coloma:2015kiu}. In particular, simulations by using data from Deep Underground Neutrino Experiment (DUNE) have allowed to obtain different estimations for the constraints on NSI parameters which were showed in Refs.~\cite{Coloma:2015kiu,deGouvea:2015ndi,Brahma:2022xld,Bakhti:2022axo}. For instance, one of the most important projections was presented in~\cite{Bakhti:2022axo}, $\epsilon^{d}_{\mu\tau} \sim 5.5\times 10^{-4}$, by assuming a detection efficiency of tau neutrinos ($\nu_{\tau}$) of the order of $30\%$ as projected by DUNE Collaboration~\cite{DUNE:2020ypp}.\medskip

In general, the bounds reported by ANTARES~\cite{ANTARES:2021crm} and IceCube~\cite{IceCube:2022ubv}, as well as the projection from DUNE~\cite{Bakhti:2022axo}, were obtained through an analysis of \textit{one NSI at one time}, unlike the \textit{marginalized over NSI} analysis, where the contribution of all NSI parameters may lead to less severe constraints. As an example of the latter, a limit of $\epsilon_{\mu\tau} < 3.5\times 10^{-2}$ was achieved in Ref.~\cite{Coloma:2015kiu} as expected sensitivity of the DUNE experiment.\medskip

The main goal of this work is to explore the NP implications of the recent constraints from ANTARES~\cite{ANTARES:2021crm} and IceCube~\cite{IceCube:2022ubv} to the NSI parameter $\epsilon^d_{\mu\tau}$. For such a purpose, we adopt a phenomenological approach by considering a simplified model where neutral-current NSI is mediated by a new neutral vector boson ($Z^\prime$) with couplings to first-family quarks and lepton flavor violating (LFV) $\mu\tau$ couplings in the lepton sector. 
In this simplified $Z^\prime$ model we will consider the case where only down-quark has the propagation NSI with neutrinos, as it was considered by ANTARES~\cite{ANTARES:2021crm} and IceCube~\cite{IceCube:2022ubv}. We study the constraints of these neutrino experiments to the parametric space of this model. We incorporate in our analysis additional bounds from $(g-2)_\mu$ anomaly, the lepton flavor universality ratio of leptonic $\tau$ decays, the LFV decay $\tau \to  \mu\rho$, ATLAS experiment limits from searches for a $Z^\prime$ boson with $\mu\tau$ couplings. Moreover, we also explore the expected sensitivity reach at DUNE experiment~\cite{Bakhti:2022axo}. In general, we will show that ANTARES and IceCube can exclude a large region of the parametric space of the NP scenario under consideration. Additionally, these neutrino experiments can exclude a similar region than ATLAS experiment, showing the potential to provide complementary information to the one obtained from direct searches at the Large Hadron Collider (LHC). \medskip


This work is organized as follows. In Sec.~\ref{model} we present a brief discussion of the main components of the simplified $Z^\prime$ model to study the NP implications from ANTARES and IceCube. We also discuss the most important constraints to which this NP scenario contributes. We perform our phenomenological analysis of the excluded parametric space in Sec.~\ref{analysis}. The concluding remarks of our work are presented in Sec.~\ref{conclusions}.

\section{Simplified $Z^\prime$ model} 
\label{model}

We take a phenomenological approach by regarding a simplified model in which the NP effects originate from an extra neutral gauge boson $Z^\prime$ of mass $M_{Z^\prime}$ with generic left-hand and right-handed couplings to leptons and quarks. The relevant Lagrangian can be expressed as
\begin{eqnarray} \label{LagZ}
\mathcal{L}_{Z^\prime} &=& \bar{\ell}  \gamma^\mu  (g^L_{\ell\ell^\prime} P_{L} + g^R_{\ell\ell^\prime} P_{R} )\ell^\prime Z^\prime_\mu + \bar{\nu}_\ell \gamma^\mu (g^L_{\ell\ell^\prime}  P_{L}) \nu_{\ell^\prime} Z^\prime_\mu \nonumber \\
&&+ \bar{q}  \gamma^\mu  (g^L_{qq^\prime} P_{L} + g^R_{qq^\prime} P_{R} ) q^\prime Z^\prime_\mu ,
\end{eqnarray}

\noindent where $g^{X}_{\ell\ell^\prime}$ and $g^X_{qq^\prime}$ ($X=L,R$) are the leptonic and quark chiral couplings, respectively, with $\ell^{(\prime)}= e,\mu,\tau$ representing the lepton flavors and $q^{(\prime)}$ labeling the generations of up-type and down-type quarks. To respect the $SU(2)_L$ gauge invariance, the couplings to the left-handed neutrinos and to left-handed charged leptons must be the same; while for the same purpose in the quark sector, the couplings of the up-type quarks must be parameterised as $g_{u u^\prime} = V_{ud}\, g_{d d^\prime}V_{u^\prime d^\prime}^\dag$.  For simplicity we will take into the quark sector the interaction to be flavor diagonal. The leptonic couplings can be flavor-preserving (diagonal) $\ell = \ell^\prime$ and/or flavor-violating (off-diagonal) $\ell \neq \ell^\prime$.  In addition, it is well known that NSI effects in propagation enter only through the vector couplings, therefore, we will consider for simplicity $g^L_{qq^\prime}=g^R_{qq^\prime}$ and $g^{L}_{\ell\ell^\prime} = g^{R}_{\ell\ell^\prime}$, in both quark and lepton sectors, respectively. We will take these flavor-dependent couplings to be real.\medskip

 Although our approach in this work will be phenomenological, a few comments about the ultra violet (UV)-complete model are in order. The underlying UV complete model giving rise to the Lagrangian of Eq. (\ref{LagZ}) must respect electroweak symmetry breaking and so forth, the interaction of fermions to the extra fields must be sizeable. The simplest extension for the SM that includes a $Z^\prime$ boson can be realized through an extra $U(1)^\prime$ gauge group. To the best of our knowledge, it is not simple in this kind of $U(1)^\prime$ models to generate flavor violating neutrino couplings, since there are strong bounds becoming from charged LFV processes, but this property can be achieved through a non trivial two component representation  of the left handed leptons, as it was pointed out in Ref.~\cite{Farzan:2015hkd,Farzan:2019xor,Farzan:2016wym}. For clarity, we will not consider here extra fields, such as leptons model dependent, needed to anomaly cancellation or any further extra field. Additionally, although baryon number is needed in order to have an interaction between quarks and the $Z^\prime$ gauge boson, opposite to other $U(1)^\prime$ realizations, the UV described above does not need right handed neutrinos for anomaly cancellation.  Thus, within this simplified $Z^\prime$ scenario, we will not include right-handed neutrinos. \medskip

From the simplified $Z^\prime$ model describes by Eq.~\eqref{LagZ} we can generate the neutral-current NSI dimension-six operator $(\bar{\nu}_\mu \gamma^\alpha P_L \nu_\tau)(\bar{d} \gamma_\alpha P_L d)$, where the neutrinos are assumed to have off-diagonal $\mu\tau$ couplings and interact with down-quark. The corresponding NSI parameter has the form
\begin{equation} \label{emutau}
\epsilon^d_{\mu\tau} = \frac{1}{2\sqrt{2}G_F} \frac{g^L_{\mu\tau} g^L_{dd}}{M_{Z^\prime}^2}.
\end{equation}

\noindent As we mentioned above, in the analyses to obtain the limits on the NSI parameter $\epsilon^{d}_{\mu\tau}$, the ANTARES~\cite{ANTARES:2021crm} and IceCube~\cite{IceCube:2022ubv} experiments considered neutrino energies in the range 16 GeV to 100 GeV, and 100 GeV to 10 TeV, respectively. Neutrinos with this kind of energies interact with the matter by \textit{elastic forward scattering} that occurs in the limit of a transferred momentum $q=0$ \cite{Coloma:2020gfv}, unlike neutrinos with energies above 10 TeV which undergo \textit{deep inelastic scattering} \cite{Giunti:2007ry,Pandey:2019apj}. 
Likewise the Mikheyev–Smirnov–Wolfenstein (MSW) effect mechanism, included in the analyses of ANTARES and IceCube, describes that neutrinos propagating in matter undergoes a potential due to the \textit{forward elastic scattering} with the nucleons and electrons in the medium\cite{Giunti:2007ry}. In this regard, we take into account a null transferred momentum ($q=0$) in the previous expression, Eq.~\eqref{emutau}.

\subsection{Constraints}

In the following, we present the various relevant processes to which our simplified $Z^\prime$ model contributes and summarize all the experimental constraints. There are additional constraints coming from LEP and electroweak data which turned out to be weak~\cite{Altmannshofer:2016brv} and therefore we will not take them into account.
 
 \subsubsection{$a_\mu = (g-2)_\mu$}
 
A $Z^\prime$ boson with LFV $\mu\tau$ couplings can contribute to the one-loop level to the observed anomalous magnetic moment of the muon, $a_\mu = (g-2)_\mu$~\cite{Altmannshofer:2016brv}. This is the so-called \textit{leptophilic $Z^\prime$ scenario} with only a ﬂavor off-diagonal coupling to the muon and tau sector.\footnote{For a recent global analysis of the phenomenology of leptophilic $Z^\prime$ bosons in all leptonic sectors, see, Ref.~\cite{Buras:2021btx}.} In the heavy limit $M_{Z^\prime} \gg m_\tau$, the dominant one-loop contribution for $g^L_{\mu\tau} = g^R_{\mu\tau}$ (vector coupling) can be written as
\begin{equation}
    \Delta a_\mu = \frac{1}{12\pi^2}\frac{m_\mu^2}{M_{Z^\prime}^2} \Big(3\frac{m_\tau}{m_\mu}  -2 \Big) \vert g^L_{\mu\tau} \vert^2,
\end{equation}

\noindent where $m_\tau$ and $m_\mu$ are the masses of the tau and muon, respectively. 
In 2021, the Muon $g-2$ experiment at FNAL confirmed~\cite{Muong-2:2021ojo} the anomaly observed by BNL in 2001~\cite{Muong-2:2006rrc}. The combined experimental value is $a_\mu^{\rm Exp} = (116592061 \pm 41)\times 10^{-11}$~\cite{Muong-2:2021ojo}. 
On the theoretical side, the SM prediction depends on the estimation of the hadronic vacuum polarization (HVP). Currently, there is a tension in the HVP evaluations between the results obtained by the data-driven method (the so-called White Paper (WP))~\cite{Aoyama:2020ynm} and lattice QCD (BMW Collaboration)~\cite{Borsanyi:2020mff}. The deviations between experimental average and the different SM estimations $\Delta a_\mu = a_\mu^{\rm Exp} - a_\mu^{\rm SM}$, are  
\begin{eqnarray}
    \Delta a_\mu^{\rm WP} &=& (251 \pm 59)\times 10^{-11} \ (4.2\sigma), \\
        \Delta a_\mu^{\rm BMW} &=& (107 \pm 70)\times 10^{-11} \ (1.5\sigma),
\end{eqnarray}
\noindent where WP calculation exhibits a higher tension than the one from BMW. The possible origin of this inconsistency is a matter of debate.  In further analysis, we will use these two different values on $\Delta a_\mu$ as an instructive exercise.

 \subsubsection{LFU ratio of leptonic $\tau$ decays}
 
In addition, the LFV $Z^\prime$ couplings leads to tree-level contributions to the lepton flavor universality (LFU) ratio of leptonic $\tau$ decays, $R_{\mu/e} = {\rm BR}(\tau \to \mu\nu_\tau\bar{\nu}_\mu) / {\rm BR}(\tau \to e\nu_\tau\bar{\nu}_e)$~\cite{Altmannshofer:2016brv}. In the vector coupling case $g^L_{\mu\tau} = g^R_{\mu\tau}$, the NP effects on $R_{\mu/e}$ can be expressed as~\cite{Altmannshofer:2016brv}
\begin{equation}\label{Rmue}
    R_{\mu/e} = R_{\mu/e}^{\rm SM} [(1+X_{\mu\tau})^2+ 3X_{\mu\tau}^2] ,
\end{equation}
where 
\begin{equation}
    X_{\mu\tau} = \frac{1}{2\sqrt{2}G_F} \frac{\vert g_{\mu\tau}^L \vert^2}{M_{Z^\prime}^2} ,
\end{equation}

\noindent and $R_{\mu/e}^{\rm SM} = 0.972559 \pm 0.000005$~\cite{Pich:2013lsa}. This theoretical value has to be compared with the experimental value reported by HFLAV of $R_{\mu/e}^{\rm Exp} = 0.9761 \pm 0.0028$~\cite{HFLAV:2019otj}. From Eq.~\eqref{Rmue} we can obtain the bound
\begin{equation}
    g^L_{\mu\tau}= \frac{\big(4\sqrt{2}G_F \big)^{1/2}}{2\sqrt{2}} M_{Z^\prime} \big[(1+ 4\Delta R_{\mu/e})^{1/2} \big]^{1/2},
\end{equation}
with
\begin{equation}
    \Delta R_{\mu/e} = \frac{R_{\mu/e}^{\rm Exp}}{R_{\mu/e}^{\rm SM}}  -1 = (3.6 \pm 2.9)\times 10^{-3} .
\end{equation}

\subsubsection{LFV hadronic $\tau$ decay $\tau \to \mu \rho$}

The product of couplings $g^L_{\mu\tau}g^L_{dd}$ generates the process $\tau \to \mu d\bar{d}$, which lead to the LFV hadronic $\tau$ decay $\tau \to \mu \rho$. By means of the hadronic matrix element $\langle 0 \vert \bar{u}\gamma^\mu u - \bar{d}\gamma^\mu d\vert \rho \rangle = i f_\rho m_\rho \epsilon^\mu_\rho$ (with $f_\rho, m_\rho,$ and $\epsilon_\rho$ being the decay constant, mass, and polarization vector of the neutral $\rho$ meson, respectively), the branching ratio of $\tau \to \mu \rho$ is given by 
\begin{eqnarray}
    {\rm BR}& & (\tau \to \mu\rho) = \nonumber \\
   && \frac{f_\rho^2 m_\tau^3}{128\pi \Gamma_\tau} \left(1+2 \frac{m_\rho^2}{m_\tau^2} \right) \left(1- \frac{m_\rho^2}{m_\tau^2} \right)^2 \frac{\vert g^L_{\mu\tau} g^L_{dd}\vert^2}{M_{Z^\prime}^4}.
\end{eqnarray}


\noindent From the experimental data for lepton-pair decay $\rho \to e^+ e^-$~\cite{PDG2022} and the theoretical expression 
\begin{equation}
{\rm BR} (\rho \to e^+ e^-) = \frac{1}{\Gamma_\rho} \frac{2\pi}{3} \frac{\alpha^2 f_\rho^2}{m_\rho},
\end{equation}
one gets the decay constant value $f_\rho = 0.221$ GeV, with very small uncertainties. 
Recently, the Belle experiment has reported an updated upper limit of the branching ratio of $\tau^- \to \mu^-\rho^0$, ${\rm BR}(\tau^- \to \mu^-\rho^0) < 1.7\times 10^{-8}$ at $90\%$ of confidence level, using the full 980 fb${}^{-1}$ data set~\cite{Belle:2023ziz}. This limit can be translated into the strong bound
\begin{equation}
\frac{\vert g^L_{\mu\tau} g^L_{dd}\vert}{M_{Z^\prime}^2}  < 7.9 \times 10^{-9}  \ {\rm GeV}^{-2}.
\end{equation}

\noindent It is expected that Belle II with a integrated luminosity of 50 ab${}^{-1}$ would be able to improve this UL down to $2\times 10^{-10}$~\cite{Belle-II:2018jsg}.

\begin{figure*}[!t]
\hspace{0.5cm} Leptophilic-like $\mu\tau$ scenario \medskip

\includegraphics[width=0.47\textwidth]{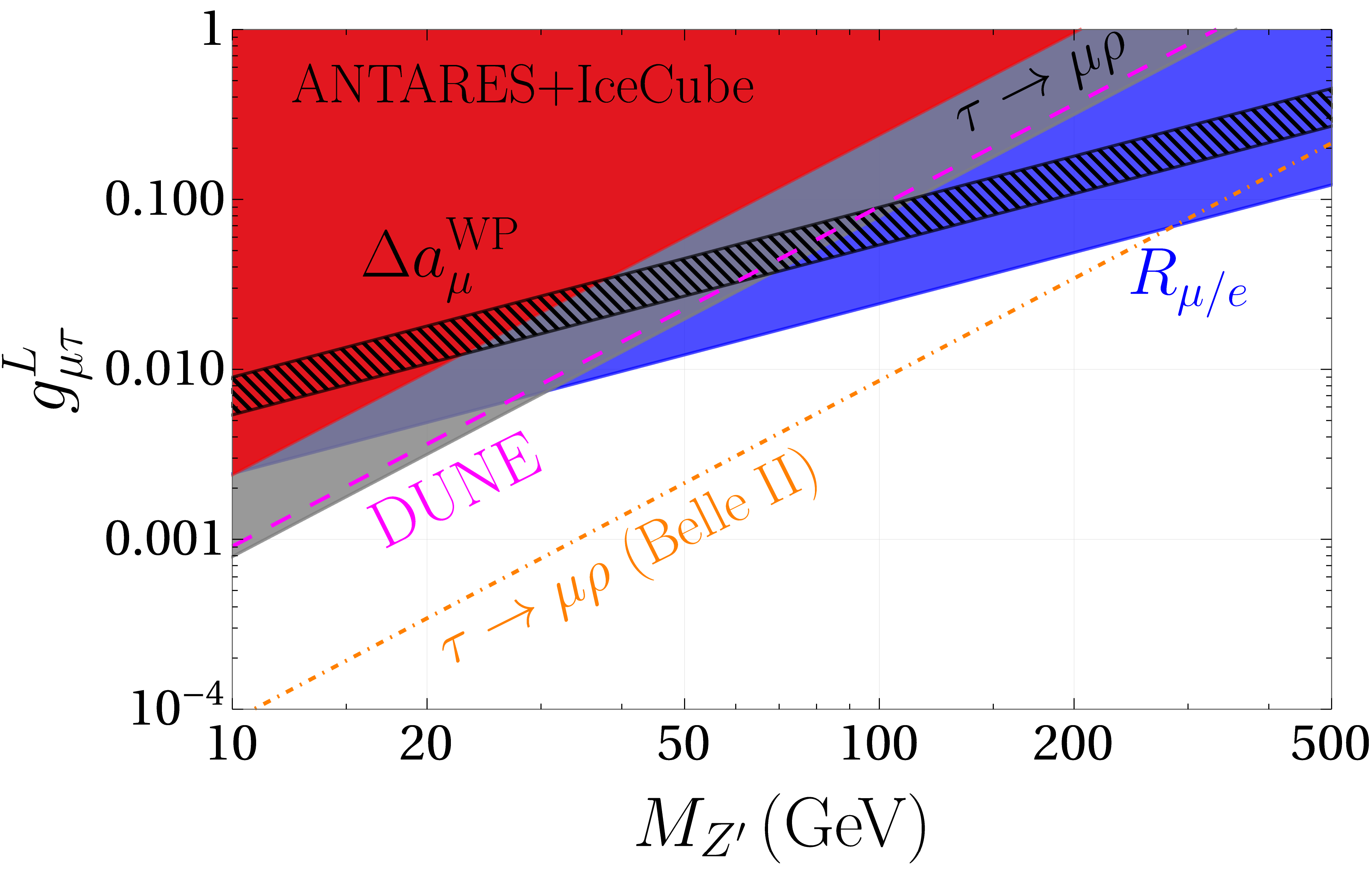} \
\includegraphics[width=0.47\textwidth]{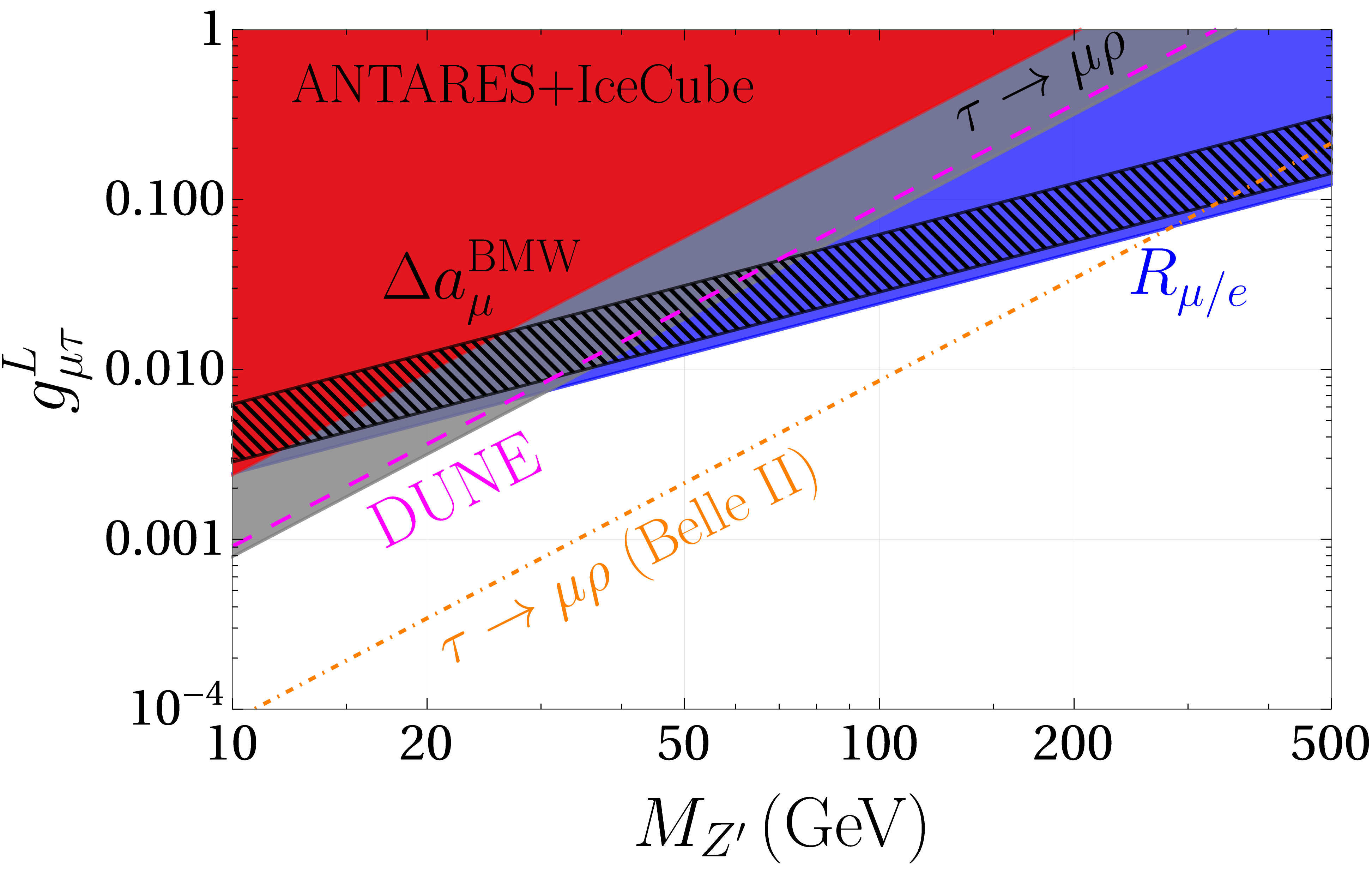} 
\caption{Parametric space $(M_{Z^\prime},g^L_{\mu\tau})$ in the $Z^\prime$ mass range $[10,500]$ GeV for the leptophilic-like $\mu\tau$ scenario. The black hatched region corresponds to the allowed parameter space to explain $\Delta a_\mu$ estimated by WP (left) and BMW (right). The blue and gray regions are excluded by the LFU ratio $R_{\mu/e}$ and $\tau\to\mu\rho$. The bound from ANTARES+IceCube~\cite{ANTARES:2021crm,IceCube:2022ubv} is represented by the red region. DUNE~\cite{Bakhti:2022axo} and Belle II~\cite{Belle-II:2018jsg} projections are shown by the magenta dashed and orange dot-dashed lines, respectively.}
\label{Fig1}
\end{figure*}

\subsubsection{LHC searches}

 Searches for a $Z^\prime$ boson with LFV couplings ($\mu\tau$) have been carried out by the ATLAS~\cite{ATLAS:2016loq,ATLAS:2018mrn} and CMS~\cite{CMS:2022fsw} experiments at the LHC in $pp$ collisions at $\sqrt{s} = 13$ TeV. These analyses were performed  through the resonant production process $pp \to Z^\prime \to \mu\tau$, by assuming that $Z^\prime$ boson has the same fermion couplings as the SM $Z$ boson in the quark sector, but allowing LFV leptonic $\mu\tau$ in the final states~\cite{ATLAS:2016loq,ATLAS:2018mrn,CMS:2022fsw}. No evidence for LFV signals was found and ULs were set (at $95\%$ confidence level) on the product of cross section and branching ratio, excluding $Z^\prime$ masses up to 3.5 TeV~\cite{,ATLAS:2018mrn} and 4.1 TeV~\cite{CMS:2022fsw}, respectively. In the case of CMS~\cite{CMS:2022fsw}, the exclusion was obtained with an assumed branching ratio of $10\%$ (${\rm BR}(Z^\prime \to \mu\tau) = 10\%$). Since the leading order production of such a LFV $\mu\tau$ signal involves first quark generation, the LHC bounds are applicable in our phenomenological approach of a simplified $Z^\prime$ boson. To be conservative, we will incorporate in our analysis the bounds obtained by ATLAS~\cite{ATLAS:2018mrn}, since the ones from CMS~\cite{CMS:2022fsw} will require additional assumptions. \medskip

We close by mentioning that a $d$-quark coupling $g^L_{dd}$ with sizeable values can generates the 4-quark interaction $(\bar{d} \gamma^\mu P_L d)(\bar{d} \gamma_\mu P_L d)$. However, this operator has no phenomenological consequences at low-energy physics, for example, pion physics. On the other hand, at high-energy, this operator can induce the process $pp \to Z^\prime \to d\bar{d}$ (di-jet). This di-jet searches has been pursued at the LHC. The CMS Collaboration has obtained 95\% confidence level lower limits on high resonance masses between 3.1 and 3.3 TeV for a $Z^\prime$ boson with SM-like couplings~\cite{CMS:2019gwf}. The mass range explored for this leptophobic $Z^\prime$ scenario is above the one studied in our work; therefore, it will be not included in our analysis.

\section{Analysis and discussion}
\label{analysis}

In this section we present the study of the allowed parametric space of the simplified $Z^\prime$ model with couplings to the down-quark ($g^L_{dd}$) and LFV $\mu\tau$ couplings in the lepton sector ($g^L_{\mu\tau}$). This model consists of the three free parameters: $M_{Z^\prime}$, $g^L_{\mu\tau}$, and $g^L_{dd}$.
We will survey the LFV $\mu\tau$ coupling in the range $[10^{-8},1]$ and study two scenarios for the down-quark coupling, that is, ($i$) \textit{leptophilic-like $\mu\tau$ scenario} in which $g^L_{dd} \ll 1 \ (\sim \mathcal{O}(10^{-3}))$, and ($ii$) \textit{non-leptophilic scenario} in which $g^L_{dd}$ has a sizeable value.
Moreover, we will pay attention to scenarios with a $Z^\prime$ boson heavier than the $\tau$ lepton ($M_{Z^\prime} \gg m_\tau$). Specifically, we will explore the parametric space in the mass range $M_{Z^\prime} \in [10, 3000]$ GeV. For a $Z^\prime$ lighter than the tau, the bound from $\tau \to \mu Z^\prime$ put very strong constraints~\cite{Altmannshofer:2016brv}; therefore, we will not study it in our analysis. \medskip

For the leptophilic-like $\mu\tau$ scenario, in Fig.~\ref{Fig1} we show the parametric space $(M_{Z^\prime},g^L_{\mu\tau})$ in the $Z^\prime$ mass range $[10,500]$ GeV, where the black hatched region corresponds to the allowed ($2\sigma$) parameter space to explain $\Delta a_\mu$ obtained by WP (left)~\cite{Aoyama:2020ynm} and BMW (right)~\cite{Borsanyi:2020mff}. 
The blue and gray regions are excluded by the LFU ratio of leptonic $\tau$ decays $R_{\mu/e}$ and the LFV decay $\tau \to  \mu\rho$. The joined bound on NSI parameter $\mu\tau$ from ANTARES~\cite{ANTARES:2021crm} and IceCube~\cite{IceCube:2022ubv} (referred to by us as ANTARES+IceCube) is shown by the red region. We will take into account the most restrictive limits on $\epsilon_{\mu\tau}^d$ in the scenario of NH.
To further discussion we have also included the projections from: 
1) DUNE (magenta dashed line) that would have the sensitivity of probing the NSI parameter $\epsilon_{\mu\tau}^d$ around $\epsilon_{\mu\tau}^d < 5.5\times 10^{-4}$~\cite{Bakhti:2022axo}. This result corresponds to an estimation by considering an optimistic $30\%$ of detection efficiency of tau neutrinos~\cite{Bakhti:2022axo}, and 
2) Belle II (orange dot-dashed line) sensitivity on $\tau \to\mu\rho$ for an integrated luminosity of 50 ab${}^{-1}$~\cite{Belle-II:2018jsg}. 

The most important remarks are the followings:
\begin{itemize}
\item We observe that the LFU ratio $R_{\mu/e}$ excludes the whole region allowed to address the most recent $\Delta a_\mu$ estimations either WP or BMW. Such a result it was previously pointed out in Ref.~\cite{Altmannshofer:2016brv} by considering the E821-BNL measurement.
\item In the mass window $M_{Z^\prime} \sim [10,20] \ {\rm GeV}$, it is observed that the joined bound from ANTARES+IceCube gives complementary constraints to the one obtained from the LFU ratio $R_{\mu/e}$. While the LFV decay $\tau \to  \mu\rho$, provides strong constraints on LFV coupling by excluding values down to $\sim 10^{-3}$.
\item Concerning to the DUNE projection~\cite{Bakhti:2022axo}, it would improve by (around) one order the magnitude the excluded region by ANTARES+IceCube up to masses of $\sim 30$ GeV. Moreover, this experiment would be able to exclude a similar region that the one obtained by the LFV decay $\tau \to  \mu\rho$.
\item In the case of the perspectives on $\tau \to  \mu\rho$ at Belle II, the excluded region would be stronger than the one from $R_{\mu/e}$, almost in the full mass region.
\item Our analysis shows that there is an intriguing synergy between neutrino experiments, such as ANTARES and IceCube (as well as DUNE in the future), and low-energy flavor experiments to put constraints to a $Z^\prime$ boson with LFV $\mu\tau$ couplings. Although our approach in this work is phenomenological, we expect that similar implications can be extrapolated to an UV complete model.
\end{itemize} \medskip

\begin{figure}[t!]
\hspace{0.8cm} Non-leptophilic scenario
\includegraphics[width=0.48\textwidth]{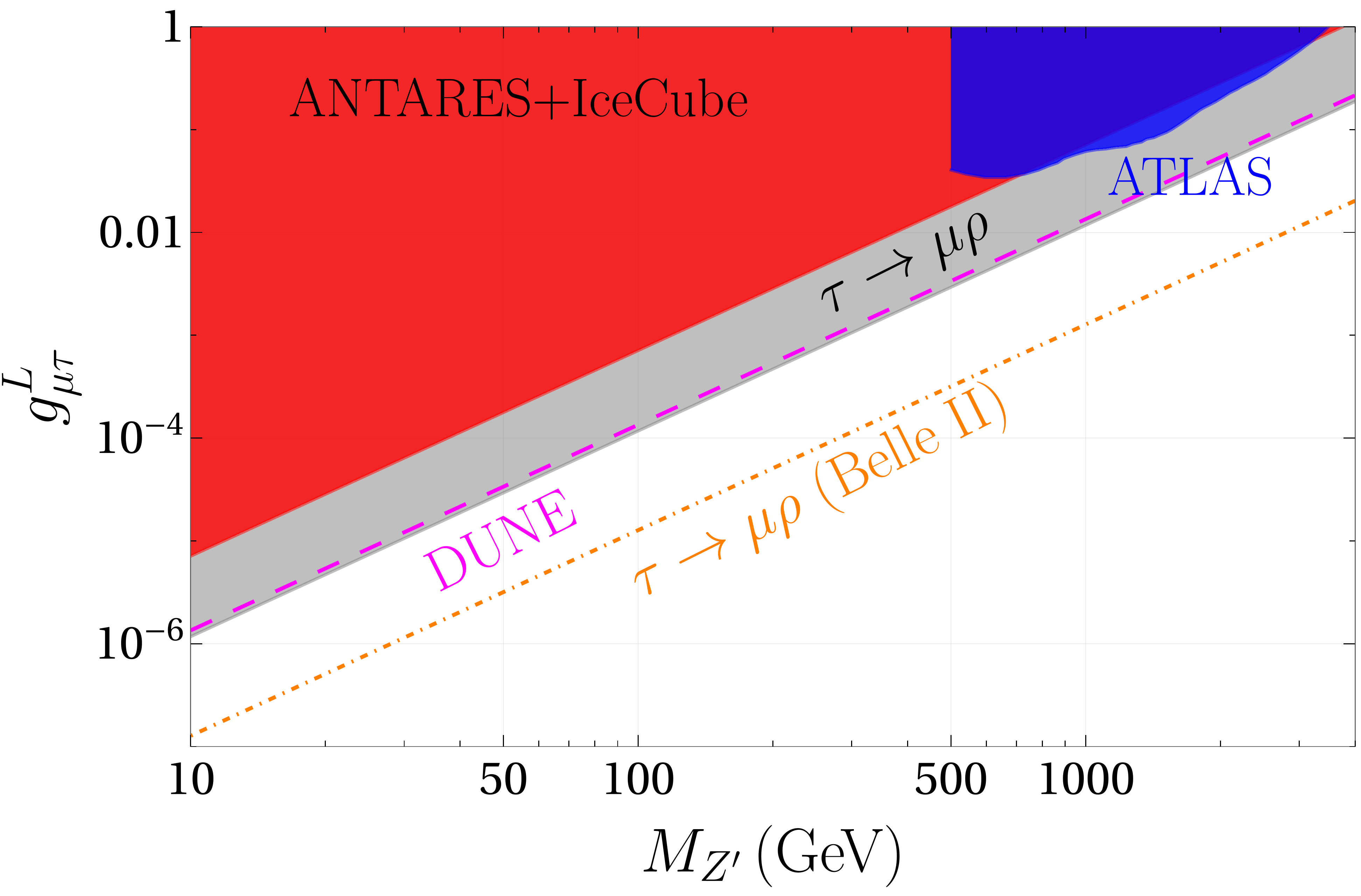}
\caption{Parametric space $(M_{Z^\prime},g^L_{\mu\tau})$ in the $Z^\prime$ mass range $[10,3000]$ GeV for the non-leptophilic scenario. It is assumed that the $Z^\prime$ boson has the same quark coupling as the SM $Z$ boson in the bounds from ANTARES+IceCube, $\tau \to\mu\rho$, and DUNE. The blue region is excluded by ATLAS~\cite{ATLAS:2018mrn}.}
\label{Fig2}
\end{figure}

For the mass range $[10, 3000]$ GeV, in Fig.~\ref{Fig2} we show the parametric space $(M_{Z^\prime},g^L_{\mu\tau})$ for the non-leptophilic scenario. For the higher mass range we have included the bounds obtained by ATLAS~\cite{ATLAS:2018mrn} (blue region). To properly compare with the ATLAS limits~\cite{ATLAS:2018mrn}, in the joined bound from ANTARES+IceCube, DUNE, and LFV decay $\tau \to  \mu\rho$, the NP quark coupling $g^L_{dd}$ must be considered to has the same quark couplings as the SM $Z$ boson ($g^L_{dd} \simeq 0.67$). It is obtained that:
\begin{itemize}
\item The strongest constraint is given by $\tau \to  \mu\rho$ in the whole parametric space.
\item The joined bound from ANTARES+IceCube excludes a similar region than ATLAS.
\item DUNE would be able to exclude a similar region that the one obtained by $\tau \to  \mu\rho$.
\item The stronger bounds are expected to be obtained by Belle II.
\end{itemize}

\noindent Interestingly, this shows that neutrino experiments such as ANTARES, IceCube, and DUNE have the potential to provide complementarity with the direct searches performed at the LHC and flavor physics. 


\section{Conclusions} 
\label{conclusions}

We have studied the NP implications of the recent constraints from ANTARES and IceCube to the NSI parameter $\epsilon^d_{\mu\tau}$. For a such purpose, from a phenomenological approach, we have implemented a simplified $Z^\prime$ boson model to generates the neutral-current NSI operator $(\bar{\nu}_\mu \gamma^\alpha P_L \nu_\tau)(\bar{d} \gamma_\alpha P_L d)$ with couplings to the ﬁrst $d$-quark generation ($g^L_{dd}$) and LFV $\mu\tau$ couplings in the lepton sector ($g^L_{\mu\tau}$). 
Within this framework, we studied the parametric space of the leptophilic-like $\mu\tau$ and non-leptophilic scenarios for a $Z^\prime$ boson with a mass heavier than the $\tau$ lepton.
To provide a comparative analysis to the joined constraint from ANTARES and IceCube (ANTARES+IceCube), we have taken into account the most important additional constraints such as, $(g-2)_\mu$ anomaly, the LFU ratio of leptonic $\tau$ decays $R_{\mu/e}$, the LFV decay $\tau \to  \mu\rho$, ATLAS limits from searches for a $Z^\prime$ boson with $\mu\tau$ couplings. The expected sensitivity reach at DUNE and Belle II (50 ab$^{-1}$) experiments are discussed as well. 
 
In leptophilic-like $\mu\tau$ scenario, we have found that for the mass window $M_{Z^\prime} \sim [10,20] \ {\rm GeV}$, ANTARES+IceCube excludes the region allowed to address the $(g-2)_\mu$ anomaly by a leptophilic $Z^\prime$ boson (in the scenario $g^L_{\mu\tau} = g^R_{\mu\tau}$). Moreover, it supplies additional bounds to the strong one obtained by the LFV decay $\tau \to  \mu\rho$. 
Improvements of approximately an order of magnitude (or higher) are expected to be achieved by the DUNE and Belle II experiments. 
Regarding the general non-leptophilic scenario, it is obtained that ANTARES+IceCube (as well as DUNE projection) put severe constraints to this scenario. Moreover, it complements the strongest constraint obtained from $\tau \to  \mu\rho$ (and its Belle II projection) in the whole parametric space. In addition, for $M_{Z^\prime} \in  [500, 3000]$ GeV, ANTARES+IceCube gives similar bound to the one obtained from the ATLAS experiment showing that such a neutrino experiments can (indirectly) supplement the NP searches performed at the LHC.

The outcomes showed in this study were obtained by using current results from ANTARES and IceCube, however, the expectations to significantly improve the sensitivity to NSI from future data of these two telescopes are quite promising. Regarding the ANTARES experiment, its next-generation successor, KM3NeT, under construction in the Mediterranean Sea, will have the sufficient potential to detect almost imperceptible effects related to NSI~\cite{HernandezRey:2021qac}. Concerning IceCube DeepCore experiment, the new extension, known as IceCube Upgrade, will allow for a significant increase in events statistics as well as the detection efficiency, and will reach lower threshold energies than DeepCore, which will also make it easier to determine very small NSI effects~\cite{IceCubeCollaboration:2021euf,IceCube:2023ins}. Consequently, our future expectations for improving the results of the present work with these coming experimental data are also encouraging.

As a final remark, our analysis showed that there is an interesting synergy between neutrino experiments, such as ANTARES and IceCube (as well as DUNE in the future), low-energy flavor experiments, and the LHC to constraining a $Z^\prime$ boson with LFV $\mu\tau$ couplings.

\acknowledgments
N.~Quintero-Poveda acknowledges financial support from Minciencias CD 82315 CT ICETEX 2021-1080. A. Parada thanks Escuela Superior de Administración Pública (ESAP) for the support through research project E5\_2023\_8.  



\end{document}